\documentclass[epj]{svjour}
\usepackage{graphics}
\usepackage{epsfig}
\usepackage{amsmath}
\usepackage{longtable}
\begin{document}
\title{Physics at ELSA}
\author{Hartmut Schmieden\inst{1} 
        \and the Crystal-Barrel/TAPS collaboration
}                     
%
%
\institute{Physikalisches Institut, 
Rheinische Friedrich-Wilhelms-Universit\"at Bonn}
\date{Received: date / Revised version: date}
%
\abstract{
Recent experimental results of the Crystal-Barrel/TAPS
collaboration at ELSA are presented. 
The experiments used a tagged photon 
beam incident on proton and neutron targets. 
Multiple photon final states were detected in the hermetically 
closed Crystal Barrel/TAPS detector setup, enabling the reconstruction of
various neutral meson production channels.
In addition to cross sections also photon beam asymmetries  
and, in the $K^0_s\Sigma^+$ channel, recoil polarisations were
determined.
\PACS{
      {13.60.-r}{Photon and charged-lepton interactions with hadrons} \and
      {13.60.Le}{Meson production} \and
      {13.88.+e}{Polarization in interactions and scattering} \and
      {14.20.Gk}{Baryon resonances with S=0}
     } 
} 
\maketitle
\section{Introduction}
\label{sec:intro}

The rich excitation spectrum of the nucleon reflects
its complicated inner dynamics.
Hence baryon spectroscopy is expected to provide
benchmark data for any model of the nucleon,
e.g. quark models in their variety \cite{CR00,Loering01} or,
increasingly in the near future,
Lattice QCD as an approximation of full Quantum Chromodynamics
\cite{KLW05}.
The major problem of nucleon spectroscopy is the width and 
density of states involved which, in many cases, prohibit their clean
identification, i.e. an unambiguous assignment of quantum numbers
within a partial wave analysis.

The current experimental base is mostly restricted to 
pion and kaon induced reactions.
However, a significant number of excited states is suspected
to have a strongly disfavoured $\pi N$ coupling \cite{CR94}.
Hence, such states might have escaped detection in conventional analyses.
This is one possible scenario why much less baryonic excitations are
observed than expected by quark models
and provided the motivation to investigate
photoinduced reactions beyond single-pion production
\cite{Crede05,SL02,Chen03}.
In particular, we studied $\eta$ photoproduction off proton and 
deuteron targets, two-meson final states, associated strangeness
photoproduction in the $K^0_s\Sigma^+$ channel, and 
$\omega$ photoproduction.

Unambiguous solutions of partial wave analyses are generally only
possible on the basis of ``complete'' experiments with regard to
the separation of reaction amplitudes.
Such measurements appear at the horizon in $\eta$ and $K$
photoproduction.
In these cases the required number of, if carefully chosen,  
8 independent observables \cite{CT97} seems accessible.
In two-meson and vectormeson photoproduction the measurement of
the required 16 or 23 quantities is presently out of range.
Nevertheless, polarisation observables provide essential 
constraints which, if not to isolate specific resonances, 
still may allow to clarify the reaction mechanism regarding
whether s-channel resonances contribute at all.


Such experiments depend on linearly and circularly polarised photon beams.
The measurement of double polarisation observables, inescapable for 
the ``complete'' experiment, require polarised nucleon targets in addition.
Eventually, the recoil nucleon polarisation needs
to be determined as well.

In the following section some of the observables of meson photoproduction
are discussed. 
Section \ref{sec:Setup} then describes the experimental setup before in
section \ref{sec:Results} some recent, 
partially preliminary, results are discussed.
The final section summarises and gives an outlook to upcoming
experiments.

\section{Cross section and polarisation observables}
\label{sec:x-sec}

With polarised beam and target, 
in the simplest case of photoproduction of single pseudoscalar
mesons the cross section can be written in the form \cite{KDT95}
\begin{eqnarray}
\label{eq:xsec}
\frac{d\sigma}{d\Omega} = \frac{d\sigma_0}{d\Omega}\:
                          [ 1 &-& P_\textrm{\tiny lin}\,\Sigma\,\cos 2\Phi \\ \nonumber
   &+& P_z\, ( -P_\textrm{\tiny lin}\, H\, \sin 2\Phi + P_\textrm{\tiny circ}\, F ) \\
                          \nonumber
   &-& P_y\, ( -T + P_\textrm{\tiny lin}\, P\, \cos 2\Phi ) \\ \nonumber
   &-& P_z\, ( -P_\textrm{\tiny lin}\, G\, \sin 2\Phi + P_\textrm{\tiny circ}\, E )
                          ],
\end{eqnarray}
where $\sigma_0$ denotes the polarisation independent cross section,
$P_\textrm{\tiny lin,circ}$ the degree of linear or circular polarisation 
of the incident photon beam,
and $\Phi$ the azimuthal orientation of the reaction plane 
with respect to the plane of linear polarisation.
The photon direction defines the $z$ axis of a right handed coordinate
frame spanned with the outgoing meson momentum $\vec k_m$:
$\vec z = \vec k_\gamma / |\vec k_\gamma|$,
$\vec y = \vec k_m / |\vec k_m|$, and
$\vec x = \vec z \times \vec y$.
$P_{x,y,z}$ are the cartesian components of the target polarisation vector
in this frame.
Once beam and/or target are polarised, the polarisation observables
$\Sigma$, $H$, $F$, $T$, $P$, $G$ and $E$ are accessible.
In particular, the measurements require the following combinations of
beam and target polarisation:
\begin{tabbing}
~~\={\bf observable}~~\={\bf beam}~~~~~~~~~~~~~~\={\bf target}\\[.1cm]
\>~~~$\Sigma$  \>linear  \>no \\
\>~~~$T$       \>no      \> transverse   \\
\>~~~$H$       \>linear  \> transverse   \\
\>~~~$F$       \>circular\> transverse   \\
\>~~~$P$       \>linear  \> transverse   \\
\>~~~$G$       \>linear  \> longitudinal \\
\>~~~$E$       \>circular\> longitudinal \\
\end{tabbing}
The polarisation observables represent ratios of 
structure functions 
and as such are related to the more general 
case of meson electroproduction.
Although more involved than photoproduction, 
through variation of the momentum transfer electroproduction is 
able to characterise the spatial structure of the processes involved
\cite{Burkert07} in addition to the time structure which is obtained through
the spectroscopic information with real photons.
If instead of, or, in addition to target polarisation 
the measurement of nucleon recoil polarisation is provided, 
further double and triple polarisation observables can be defined
which, with their mutual interrelations,
are discussed in detail in \cite{KDT95}.

In the sense of the ``complete'' experiment mentioned in the introduction,
it is necessary to determine angular distributions of at least 8 quantities.
Those must encompass the differential cross section and the 
single polarisation observables $\Sigma$ and $T$, as well as the
recoil polarisation.
Furthermore, 4 double polarisation, e.g. beam-target, observables
need to be chosen such that the occurence of discrete ambiguities can be 
avoided in the analysis of the bilinear forms of the reaction amplitudes
\cite{CT97}.

The situation becomes more complicated in the photoproduction of 
double pseudoscalars \cite{Roberts05} and vector mesons. 
Despite the infeasibility of ``complete'' experiments
with current techniques in such cases,
essential information towards the involved reaction mechanisms, 
in particular the role of s-channels resonances, 
can be expected from the investigation of single and double 
polarisation observables.

The understanding of the reaction mechanisms is a prerequisite 
for partial wave analyses (PWA) or dynamical models to disentangle
the broad and overlapping states.
Different final states and concurring (coupled) channels need
to be investigated as well as 
proton and neutron targets.
In addition to polarisation, also full coverage of the angular 
and energy range is essential. 
Thus, the combination of the Crystal Barrel \cite{CBarrel} 
and TAPS \cite{TAPS} detectors to an almost 
4$\pi$ array at the Bonn electron stretcher accelerator 
ELSA \cite{Hillert06} provides an ideal tool for the
sketched investigations.

\section{Experimental setup}
\label{sec:Setup}

The experiments were performed at the tagged photon beam of
ELSA. 
Electron beams up to $E_0 = 3.5$ GeV were used to produce
unpolarised or linearly polarised \cite{Elsner07,Elsner07a}
bremsstrahlung.
Electrons which radiated a photon were momentum analysed in
a tagging spectrometer, enabling the event-wise assignment
of the photon energy in the range $E_\gamma = 0.18 ... 0.92\,E_0$.
Photon fluxes of about $2 \times 10^7$ s$^{-1}$ were
usually used. 

\begin{figure}
\begin{center}
\resizebox{0.97\columnwidth}{!}{%
  \includegraphics{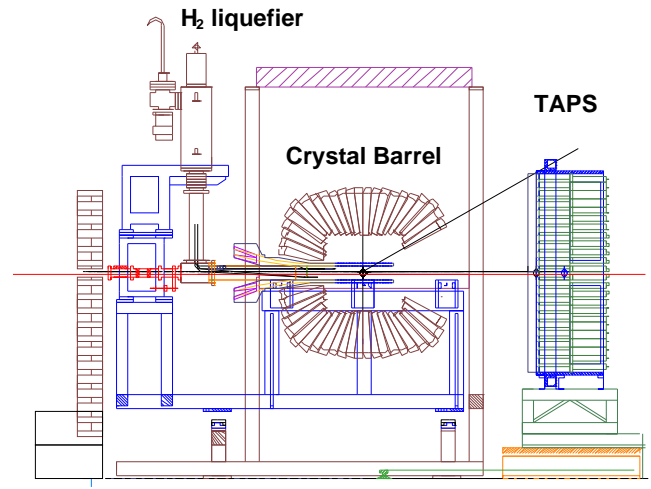} }
\end{center}
\caption{Setup of the detector system as described in the text.
         The photon beam enters from left.}
\label{fig:setup}       
\end{figure}

The detector setup is depicted in Figure\,\ref{fig:setup}.
The photon beam hit a $5.3$ cm long 
liquid hydrogen or deuterium target. 
A three layer scintillating fibre detector \cite{Suft05} ,
which surrounded the target within the polar angular range from 
15 -- 165 degrees,
was used to determine a point-coordinate for charged particles.

Both, charged particles and photons were detected in the 
Crystal Barrel detector \cite{CBarrel}. 
Its 1290 individual CsI(Tl) crystals were 
cylindrically arranged around the target in 23 rings, 
covering a polar angular range of 30 -- 168 degrees.
For photons an energy resolution of 
$\sigma_{E_\gamma}/E_\gamma 
=\linebreak 2.5\,\%/^4\sqrt{E_\gamma/\text{GeV}}$
and an angular resolution of $\sigma_{\Theta,\Phi} \simeq 1.1$\,degree 
was obtained.

The $5.8$ -- 30 degree forward cone was covered by the 
TAPS detector \cite{TAPS},
set up in one hexagonally shaped wall of 528 BaF$_2$ modules
at a distance of $118.7$\,cm from the target.
For photons between 45 and 790 MeV an energy resolution of
$\sigma_{E_\gamma}/E_\gamma 
= \left(0.59/\sqrt{E_\gamma/\text{GeV}}+1.9\right)\%$
was achieved \cite{Gabler94}.
The position of photon incidence could be resolved within
20\,mm.
For charged particle recognition
each TAPS module had a 5\,mm plastic scintillator in front of it.

\section{Recent Results}
\label{sec:Results}

The combined Crystal Barrel and TAPS setup is ideally suited 
for final states of multiple photons.
Hence, neutral mesons are very efficiently detected.
Those are of particular interest for nucleon spectroscopy 
since t-channel exchanges, which may hide resonance excitations,
are suppressed in many cases.
Investigated channels involve 
$\pi^0 \rightarrow \gamma\,\gamma$,
$\eta  \rightarrow \gamma\,\gamma$ or $3\pi^0$,
$K^0_s \rightarrow \pi^0\,\pi^0 \rightarrow 4\,\gamma$, and
$\omega\rightarrow \pi^0\,\gamma \rightarrow 3\,\gamma$,
as well as combinations thereof.
Typical invariant mass resolutions achieved were
$\sigma_{\pi^0} = 10$ MeV and $\sigma_{\eta} = 22$ MeV
in the two-photon decays of $\pi^0$ and $\eta$, and
$\sigma_{\eta} = 25$ MeV in $\eta \rightarrow 3\pi^0$.

%

\subsection{$\mathbf{\eta}$ photoproduction off the proton}
\label{subsec:eta-p}

Similar to pion photoproduction by the $\Delta(1232)$ resonance
so is $\eta$ photoproduction in the threshold region dominated by
a single nucleon resonance, the $S_{11}(1535)$.
Vice versa, $\eta$ photoproduction gives very selective access to this state,
the internal structure of which is still under debate.
Furthermore, 
$\eta$ photoproduction is a predestined channel to look for 
possible resonances with small $\pi$-N coupling which might have escaped
experimental detection to date. 
Due to its isoscalarity, the $\eta$ only connects 
$N^*$ as opposed to $\Delta^*$ resonances with the nucleon ground state, 
which simplifies the observed spectrum considerably. 

The role of resonances beyond the $S_{11}$ 
is, due to its dominance, much less clear.
The strategy to investigate this at ELSA was threefold:
First, differential cross sections have been determined over 
the full angular range across the full nucleon resonance region.
Second, as a first step towards the ``complete'' experiment
the photon beam asymmetry has been measured at the high-energy tail
of the $S_{11}$.
And third, neutron targets have been included in the investigations
as will be discussed later.

\begin{figure}
\begin{center}
\resizebox{0.85\columnwidth}{!}{%
  \includegraphics{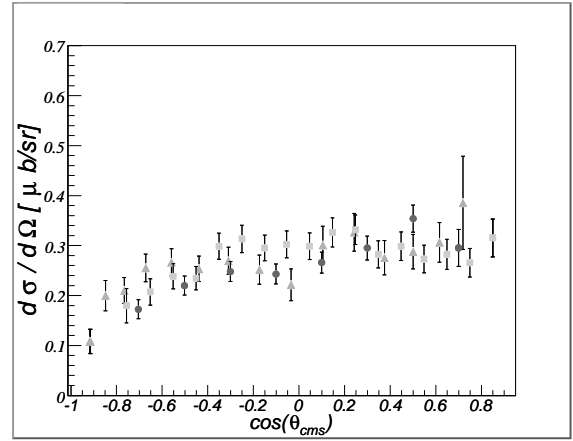} }
\end{center}
\caption{ Preliminary differential cross section with absolute
          normalisation of Crystal Barrel/TAPS \cite{Suele07} (triangles) 
          in the bin $E_\gamma=(1075 \pm 25)$\,MeV
          in comparison to earlier
          measurements from Crystal Barrel \cite{Bartholomy07} (squares)
          and CLAS \cite{Dugger02} (circles) which are normalised relative
          to the SAID $\pi^0$ photoproduction database.
        }
\label{fig:abs-x-sec}       
\end{figure}

A recent cross section measurement of the Crystal Barrel Collaboration 
at ELSA provided some indication for a yet unobserved $N^*(2070)D_{15}$
state \cite{Crede05,Bartholomy07}.
The data agree well with the CLAS data which are available at 
somewhat lower energy \cite{Dugger02}, but both data sets 
are normalised only relative to the SAID $\pi^0$ cross section.
This deficiency could now be remedied by a detailed analysis of
the photon flux based on the timing and rate information of the 
tagging system \cite{Suele07}.
The preliminary result for one energy bin is shown in Figure
\ref{fig:abs-x-sec}.
In addition to the absolute normalisation,
the full data set, which presently is in the final analysis stage, 
has also improved angular coverage in forward direction as well as 
improved statistics compared to the published data
of ref. \cite{Bartholomy07}.

Using a linearly polarised photon beam, the photon beam asymmetry $\Sigma$
was measured \cite{Elsner07,Elsner07a}
based on the azimuthal modulation of the cross section 
(cf. Eq.\,\ref{eq:xsec}).
The results agree well with the measurements at GRAAL 
\cite{Kouznetsov02,Bartalini07}.
In the light of the introductory discussion the measurement of $\Sigma$
is absolutely indispensable to disentangle the resonances possibly 
contributing to the reaction.
The analysis of the new Crystal Barrel data within the
MAID isobar model \cite{CYTD02} and the Bonn-Gatchina PWA \cite{Anisovich05}
showed, however, that no unambiguous conclusion can yet be drawn.
Both MAID and the 
Bonn-Gatchina partial wave analysis provide a satisfactory
overall description of our data.
In detail, however, there are marked differences with regard to 
the role of individual resonance contributions, cf. 
the figures and detailed discussion in \cite{Elsner07,Elsner07a}.

To resolve this problem, further double polarisation experiments are
indispensable.

\subsection{$\mathbf{\eta}$ photoproduction off the neutron}
\label{subsec:eta-n}

Off the neutron, $\eta$ photoproduction had attracted a lot of
activity recently. 
The reason is a rather narrow structure which seemed to
exhibit itself in the
cross section \cite{Kouznetsov07}. 
\begin{figure}
\begin{center}
\resizebox{0.90\columnwidth}{!}{%
  \includegraphics{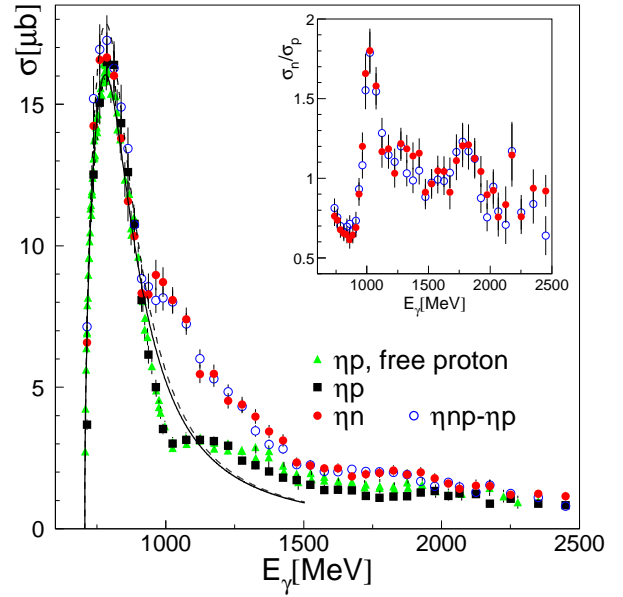} }
\end{center}
\caption{ Preliminary result for the total cross sections of
          $\eta$ photoproduction off the deuteron.
          Squares denote the quasifree $\eta$ production off the
          bound proton, filled circles off the neutron.
          The Fermi smearing has been unfolded.
          The open circles represent the difference between 
          total quasifree production off the deuteron and
          the elementary production of the free proton.
          The latter is individually shown by the triangles.
          The insert depicts the ratio of neutron over
          proton cross section.
        }
\label{fig:eta-n_x-sec}       
\end{figure}
Also in the preliminary Crystal Barrel/TAPS data which are
presented in Figure\,\ref{fig:eta-n_x-sec} there is a clear
structure visible in the neutron cross section at around
$E_\gamma \simeq 1100$ MeV \cite{Jaegle07}. 
Such a bump is present neither in the free nor the quasifree 
proton cross section.
This had triggered speculations about the influence of
a narrow (antidecuplet) state
\cite{FTP07}.
However, conventional explanations are also possible,
albeit the nature of the structure still remains under 
debate.

Interferences within the $S_{11}$ partial waves,
e.g. between $S_{11}(1535)$ and $S_{11}(1650)$,
seem to play a crucial role, both within the Bonn-Gatchina PWA
\cite{PWA-NSTAR07} and the Giessen coupled channels model 
\cite{Shklyar07}.
Interferences between different partial waves,
$S_{11}(1535)$ and $D_{13}(1520)$ are under discussion,
do exhibit in the angular distributions, 
but those can not affect the {\em total} cross section.
The direct contribution of the $D_{13}(1520)$ is much too small
to become responsible for the bump. 
This may be different for the $D_{15}(1675)$ state. 
Within the MAID model it has a particularly strong coupling to the 
neutron, at the same time however also
an unusal nucleon-$\eta$ decay branch of 17\,\%,
in disagreement with the PDG value of 0 -- 1\,\%.
To clarify this still unsatisfactory situation, measurements of
$\Sigma$, $E$ and $G$ 
are planned \cite{Krusche05a}.

\subsection{$\mathbf{K^0_s\Sigma^+}$ photoproduction}
\label{subsec:K-Sigma}

There are substantial decay branches of baryon resonances expected
into $K$-hyperon channels \cite{BIL97,CR98}.
Hence, associated strangeness photoproduction is also a channel
where so far unoberved states of the quark models may be found.
The reaction mechanism should become disentangled once
the complete experiment is approached with regard to the
reaction amplitudes.
To achieve this, the $K\,\Lambda$/$\Sigma$ channels are very well
suited due to the self analysing weak decay of the hyperons.
This facilitates the indispensable measurement of the recoil polarisation.

The $K^0\,\Sigma^+$ channel has been investigated with 
Crystal Barrel/TAPS from threshold to $W=2.3$\,GeV. 
In a first step cross sections and recoil polarisation
were determined \cite{Castelijns07} and compared to calculations within the
Bonn-Gatchina PWA \cite{Anisovich05,Sarantsev05} 
and a coupled-channels K-matrix formalism of Usov and Scholten \cite{US05}.
The total cross section is well reproduced by both calculations.
However, an additional new state at about 1840\,MeV is required
which in the PWA analysis has $P_{11}$ quantum numbers, 
in contrast to $P_{13}$ in the Usov-Scholten model.
Such an ambiguity is no surprise given the ``incompleteness'' of the
present experiments.
It will be remedied with the availability of polarisation observables.

\begin{figure*}
\begin{center}
\resizebox{0.95\textwidth}{!}{%
  \includegraphics{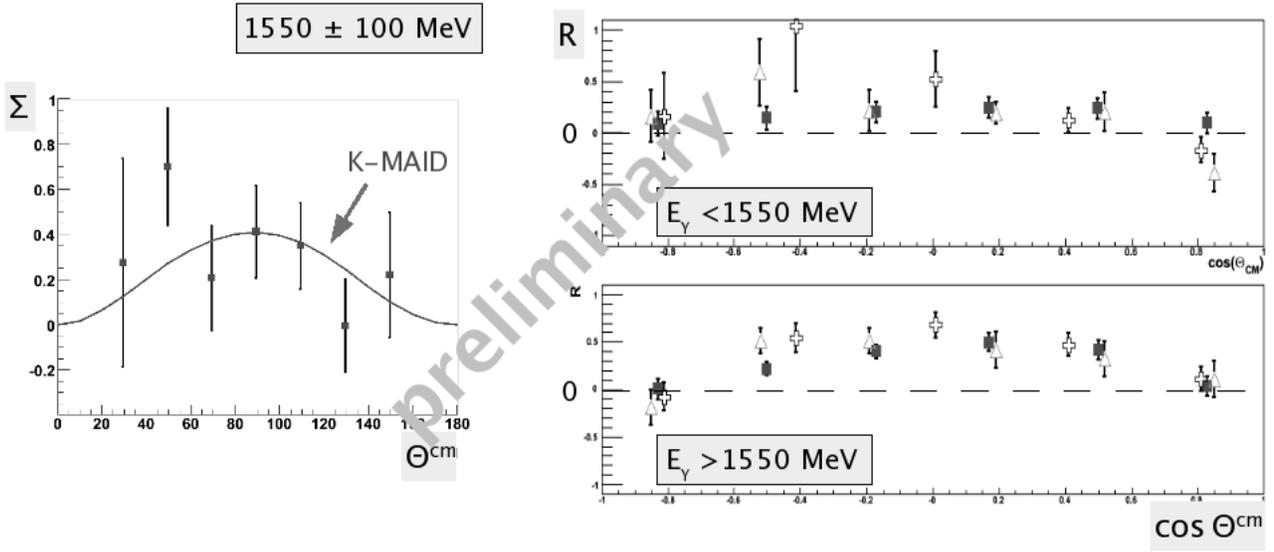} }
\end{center}
\caption{Preliminary polarisation results in $K^0_s\Sigma^+$
         photoproduction off the proton \cite{Ewald07}. 
         {\em Left:}  Photon beam asymmetry as a function of the 
                      $K$ cm-angle in the energy interval 1450--1650\,MeV.
         The errors correspond to the statistics of about a quarter
         of the full data set. 
         The curve represents the K-MAID calculation.
         {\em Right:} Recoil hyperon polarisation in two energy intervals.
         The preliminary data \cite{Ewald07} (squares) are compared to
         the previous measurements of Crystal Barrel/TAPS 
         \cite{Castelijns07} (triangles) 
         and SAPHIR \cite{Lawall05} (crosses). 
        }
\label{fig:K-Sigma}       
\end{figure*}

The first measurement of the recoil hyperon polarisation with
Crystal Barrel/TAPS \cite{Castelijns07} has been statistically 
improved in the meanwhile and, in addition, the photon beam 
asymmetry has been measured \cite{Ewald07}.
Preliminary results are shown in Figure\,\ref{fig:K-Sigma}.

\subsection{Two pseudoscalar meson channels}
\label{subsec:TwoMeson}

Double meson final states provide a tool
to get access to high-lying excitations which decay 
sequentially.
Data of $\pi^0\,\pi^0$ and $\pi^0\,\eta$  final states were
taken over the entire resonance region 
\cite{Gutz07,Sokhoyan07,Gutz07a}.

From the $2\pi^0$ channel
partial decay widths of $N^*$ and $\Delta^*$ decaying into
$\Delta(1232)\,\pi$, $N(\pi\pi)_s$, $P_{11}(1440)\pi$ and 
$D_{13}\-(1520)\pi$ were determined \cite{Thoma07}.
The PWA is compatible with known resonance properties.
The largest difference to the PDG value is found for
the total decay width of the $P_{13}(1720)$. 
From the analysis of the Roper $P_{11}$ partial wave
a strong coupling of the $N(1440)$ to $N\sigma$ is suggested
\cite{Sarantsev07}.

Due to the isoscalar $\eta$, the $\pi^0\eta$ final state provides 
selectivity to $\Delta^*$ intermediate states. 
Based on the unpolarised data, the Bonn-Gatchina PWA finds a 
$D_{33}$ partial wave with
contributions from $\Delta(1700)$ and $\Delta(1940)$ \cite{Horn07}.
Possibly, also a further $\Delta(2350)$ adds weakly.
The $\Delta(1940)$ is a highly interesting negative parity state.
In quark models it would be interpreted as a radial excitation of
the $\Delta(1700)$.
However, the mass is unexpectedly low by about 200\,MeV.

New measurements with linearly polarised photon beam
show clear beam aymmetries in both the $\pi^0\pi^0$ and
$\pi^0\eta$ channels \cite{Gutz07,Sokhoyan07,Gutz07a}.
The cross section for the production of two pseudoscalar mesons
can be written in the form
\begin{equation}
\frac{d\sigma}{d\Omega} =  \frac{d\sigma_0}{d\Omega} \,\,
\{ 1 - P_\text{lin} \left(\Sigma\,\cos 2\Phi + I^S\,\sin 2\Phi \right) \}.
\label{eq:xsec-2}
\end{equation}
Due to the 3-body final state, 
compared to Eq.\,\ref{eq:xsec}
an additional asymmtry $I^S$ occurs
which exhibits through a $\sin 2\Phi$ azimuthal modulation.
This is however found to be compatible with zero in both final states.
The $2\pi^0$ asymmetries agree well with GRAAL data published 
for the lower energies \cite{Assafiri03}.

\subsection{$\mathbf{\omega}$ photoproduction}
\label{subsec:omega}

Previous near-threshold measurements of $\omega$ photoproduction 
\cite{Barth03} had been interpreted in terms of s-channel
resonances \cite{Giessen-omega}. 
However, to disentangle resonance contributions from the 
important t-channel exchange of $\pi^0$ and/or pomeron,
polarisation observables are essential \cite{HS05}.
This motivated us to investigate the beam asymmetries
in $\omega$ photoproduction using the 
$\omega \rightarrow \pi^0\gamma$ decay.
In addition to the ordinary $\Sigma$ of Eq.\,1,
in the $\pi^0\gamma$ decay also the decay asymmetry $\Sigma_\pi$
can be accessed, which is related to the azimuthal modulation 
of the direction of the {\em decay pion} 
with respect to the photon polarisation plane.
Various models predict large negative $\Sigma$'s if s-channel
resonances contribute \cite{omega-models}. 
In this case the Bonn-Gatchina PWA expects $\Sigma_\pi \simeq 0$.
In contrast, pure t-channel mechanisms would generate
$\Sigma_\pi = \pm 0.5$. 

\begin{figure}
\begin{center}
\resizebox{0.85\columnwidth}{!}{%
  \includegraphics{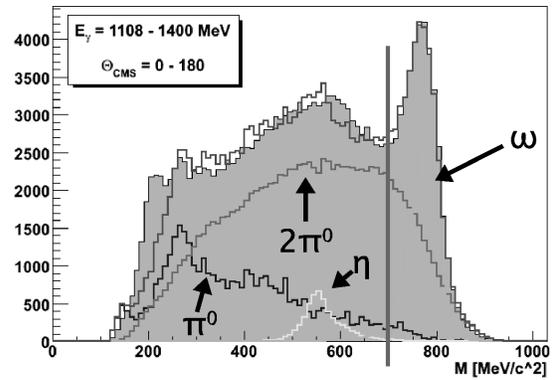} }
\end{center}
\caption{ $\pi^0\gamma$ invariant mass distribution (shaded histogram).
          As a result of Monte Carlo studies, 
          the lines show the $\omega$ signal peak and 
          main background contributions as indicated.
          The vertical line represents the minimum invariant mass
          required in the extraction of the asymmetries \cite{Klein07}. 
        }
\label{fig:omega-invmass}       
\end{figure}

Experimentally the $\omega$ is clearly identified 
in its neutral decay yielding $3\gamma$. 
However, even with the close to $4\pi$ coverage,
a number of background channels contribute significantly,
mostly due to missing one photon of competing reactions 
with 4 final state photons, or having a split-off from a 
2 photon final state. 
The shaded histogram of Figure\,\ref{fig:omega-invmass} shows the 
$\pi^0\gamma$ invariant mass distribution. 
The indicated background channels yield a reasonable
Monte-Carlo description of the spectrum.
In the region of the $\omega$ peak the far dominating part of the
background is from $2\pi^0$.
Therefore, to subtract the background, so far bin-wise fits have been
performed of signal plus $2\pi^0$ debris to the experimental distributions.
From the corresponding event numbers
within $\pm 2.5\,\sigma$ wide cuts around the peak 
the azimuthal distributions and the asymmetries were determined.

\begin{figure}
\begin{center}
\resizebox{0.85\columnwidth}{!}{%
  \includegraphics{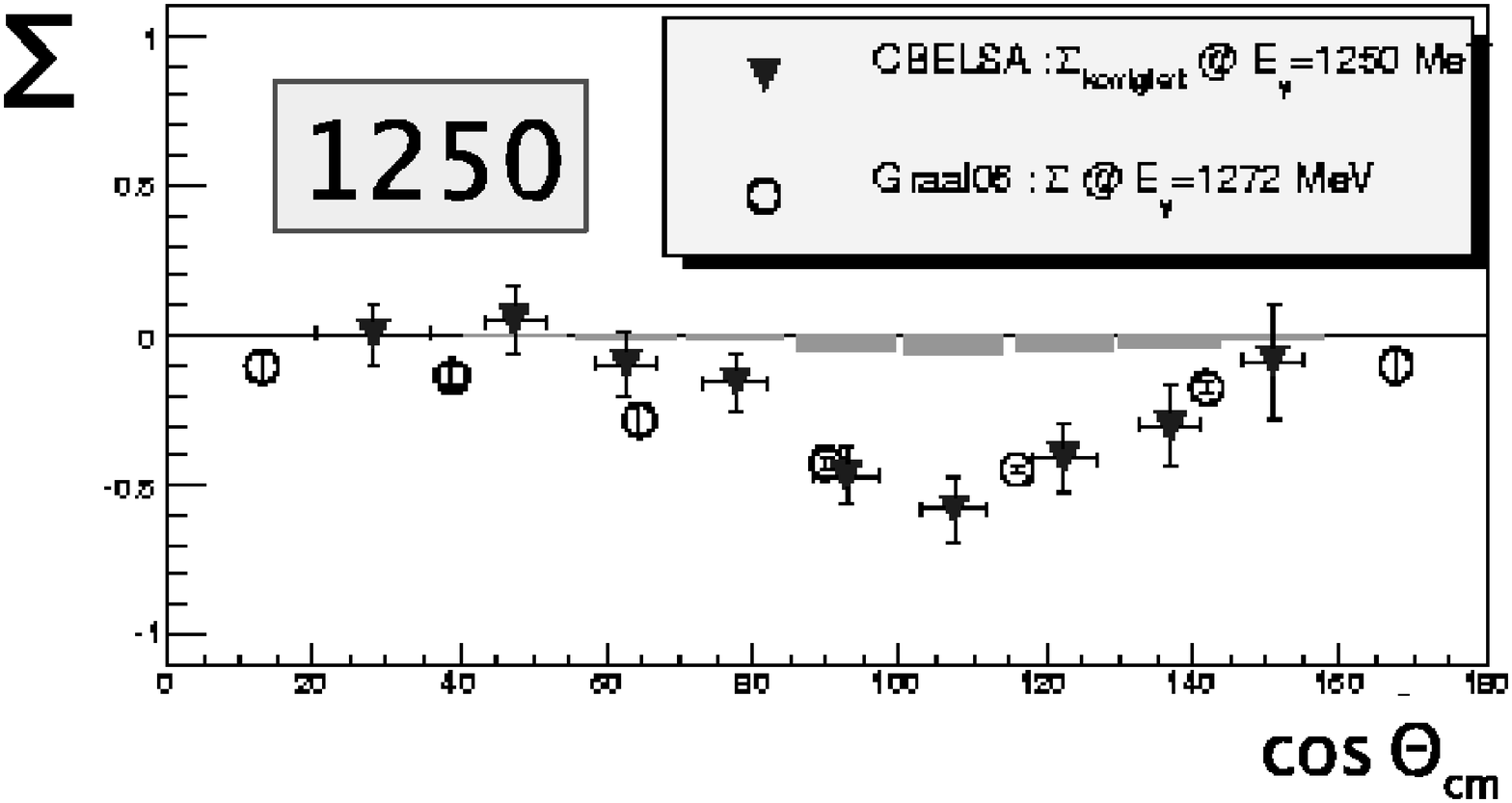} }
\resizebox{0.85\columnwidth}{!}{%
  \includegraphics{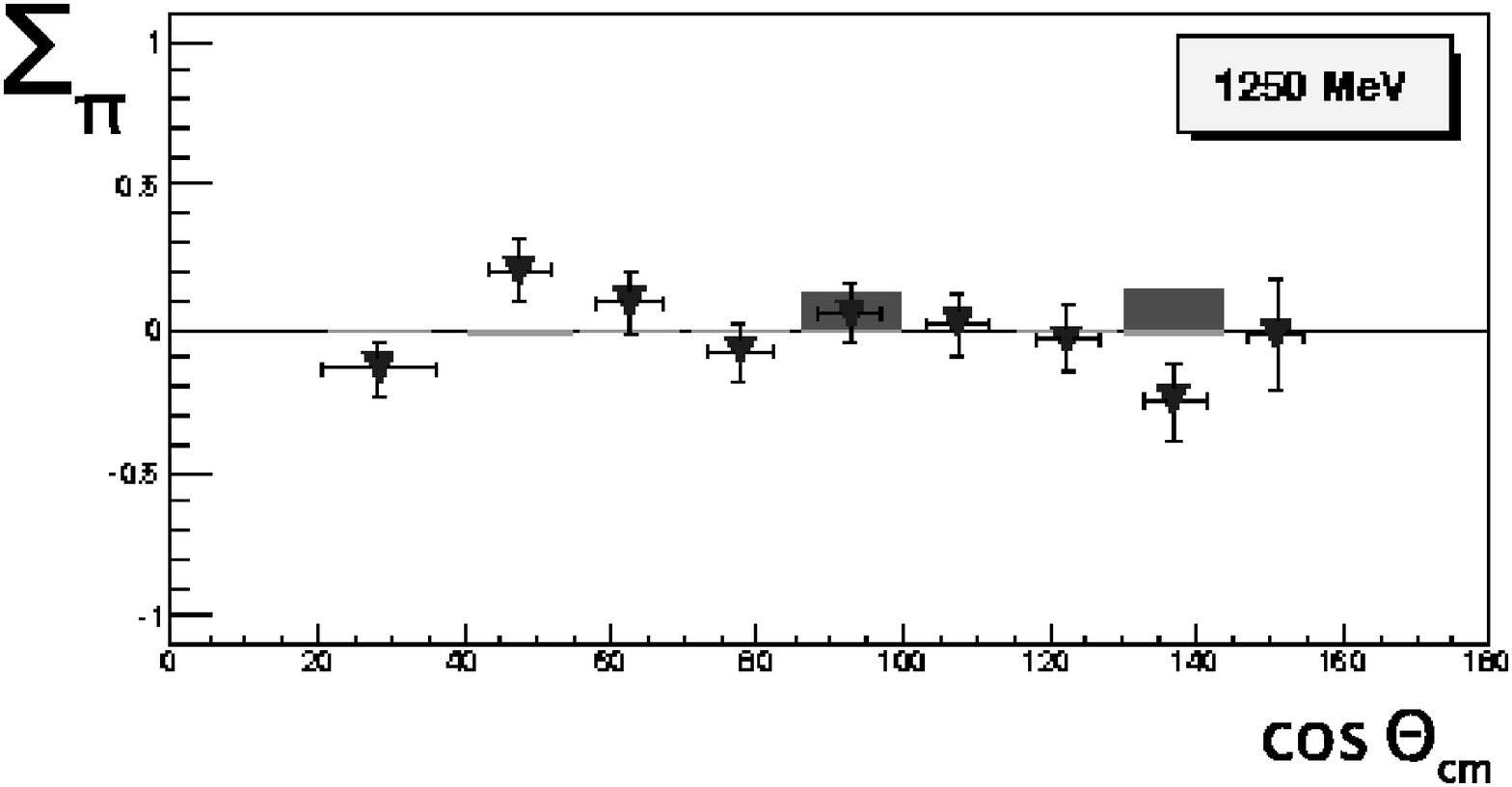} }
\end{center}
\caption{ Preliminary results (filled triangles) of the beam asymmetry 
          (upper diagram), $\Sigma$,
          and the decay asymmetry, $\Sigma_\pi$ (lower diagram)
          for the energy bin $E_\gamma=1250\pm100$\,MeV.
          GRAAL data (open circles) of $\Sigma$ are shown for comparison.
          Purely statistical errors are attached to the data points,
          the bars represent estimates of the systematic uncertainties.
        }
\label{fig:omega-Sigma}       
\end{figure}

As examples, preliminary results for $\Sigma$ and $\Sigma_\pi$
are shown in Fig.\ref{fig:omega-Sigma} in the bin of 
$E_\gamma=(1250 \pm 100)$ MeV. 
Good agreement is found to a 
recent measurement of $\Sigma$ from GRAAL \cite{Ajaka06}.
Compared to GRAAL the ELSA dataset extends the energy range to 
$E_\gamma = 1700$ MeV. 
$\Sigma_\pi$ (Fig. \ref{fig:omega-Sigma} bottom) 
is only accessible in the $\pi^0\gamma$ decay channel measured at ELSA.
Both the large negative $\Sigma$ and the small $\Sigma_\pi$ 
seem to further support that s-channel resonances provide a 
substantial contribution to $\omega$ photoproduction.
However, a closer characterisation of individual contributions appears
premature without the measurement of further double polarisation
observables, which has just started at ELSA \cite{HS05}. 

\section{Summary and outlook} 

Using the tagged photon beam of ELSA and the combined 
Crystal Barrel/TAPS setup several neutral meson
photoproduction channels have been investigated.
From the differential and total cross sections
indeed indications have been found for previously unseen 
states.
Before however unambiguous conclusions from PWA or
dynamical models are possible,
single and double polarisation observables
need to be measured as well.
As a first step, photon beam asymmtries and, in the 
$K^0_s\Sigma^+$ channel, recoil polarisations have
been determined to further constrain the analyses. 
Double polarisation experiments 
have been prepared at ELSA and are
currently underway, using the longitudinally polarised
electron beam in combination with the longitudinally
polarised frozen spin target.
Similar plans are pursued at several other laboratories. 
The double polarisation experiments
are expected to nail down the question of the existence
of at least some of the newly found states.

I am grateful to the doctoral students who provided their results
for my talk prior to publication. 
In particular I like to thank 
Kathrin Fornet-Ponse, Holger Eberhardt, Susanne Kammer,
Andre S\"ule, Frank Klein, Ralf Ewald, Ralph Castelijns, Eric Gutz,
Michael Fuchs, Vahe Sokhoyan, Igal J\"agle.
The discussions with my colleagues at the 
{\em Helmholtz Institute f\"ur Strahlen- und Kernphysik} and at the
{\em Physikalisches Institut},
B. Bantes, R. Beck, D. Elsner, J. Hannappel, 
V. Kleber, Fritz Klein, E. Klempt, and U. Thoma, 
provide a constant source of inspiration for me.
I also participated a lot of the expertise of B.\linebreak Krusche from Basel.
The enthusiasm of the accelerator group
was benefitial to all experiments; a special thank goes to W. Hillert.
Unthinkable without H. Dutz are the target polarisation experiments. 

The investigations at ELSA are supported by the federal state
of {\em North-Rhine Westphalia} and by the 
{\em Deutsche Forschungsgemeinschaft} within the SFB/TR-16.

%
%
%

\end{document}